\documentclass[%
 reprint,
superscriptaddress,
nofootinbib,
nolongbibliography,
amsmath,amssymb,
prl
]{revtex4-2}

\usepackage{comment}
\usepackage{txfonts}
\usepackage{bm,amsfonts,amsmath,amssymb,tabularx,multirow,booktabs,array,mathtools}
\usepackage{natbib}
\usepackage{xcolor}
\usepackage[dvipsnames]{xcolor}
\usepackage{textgreek}
\usepackage[utf8]{inputenc}
\usepackage[english]{babel}
\usepackage{hyperref}
\hypersetup{
    colorlinks=true,
    linkcolor=blue,
    filecolor=blue,      
    citecolor=blue,
}
\usepackage[nameinlink,capitalise]{cleveref}
\usepackage{orcidlink}
\usepackage{cancel,ulem}
\newcommand{\fnl}{f_{\rm NL}}
\newcommand{\mc}{m_{\rm c}}
\newcommand{\ms}{m_{\rm s}}

\newcommand{\ka}{\bm {k}_1}
\newcommand{\kb}{\bm {k}_2}
\newcommand{\kc}{\bm {k}_3}
\newcommand{\kab}{\bm {k}_{12}}
\newcommand{\kabc}{\bm {k}_{123}}
\newcommand{\ko}{\mathcal{Z}}

\newcommand{\cH}{\mathcal{H}}

\DeclareMathOperator{\snr}{SNR}
\DeclareMathOperator{\var}{Var}
\DeclareMathOperator{\cov}{Cov}

\defcitealias{}{}

\let\Gamma\varGamma
\let\Delta\varDelta
\let\Theta\varTheta
\let\Lambda\varLambda
\let\Xi\varXi
\let\Pi\varPi
\let\Upsilon\varUpsilon
\let\Phi\varPhi
\let\Psi\varPsi
\let\Omega\varOmega

\begin{document}

\title{Halving and Doubling: Boosting the Detection of Relativistic Effects\\ in the Galaxy Bispectrum with Optimal Subsample Selection}

\author{S.J.\ Rossiter}
\email{samanthajosephine.rossiter@unito.it}
\affiliation{Dipartimento di Fisica, Universit\`a degli Studi di Torino, Via P.\ Giuria 1, 10125 Torino, Italy}
\affiliation{INFN -- Istituto Nazionale di Fisica Nucleare, Sezione di Torino, Via P.\ Giuria 1, 10125 Torino, Italy}

\author{S.\ Camera}
\email{stefano.camera@unito.it}
\affiliation{Dipartimento di Fisica, Universit\`a degli Studi di Torino, Via P.\ Giuria 1, 10125 Torino, Italy}
\affiliation{INFN -- Istituto Nazionale di Fisica Nucleare, Sezione di Torino, Via P.\ Giuria 1, 10125 Torino, Italy}
\affiliation{INAF -- Istituto Nazionale di Astrofisica, Osservatorio Astrofisico di Torino, Strada Osservatorio 20, 10025 Pino Torinese, Italy}
\affiliation{Department of Physics \& Astronomy, University of the Western Cape, Cape Town 7535, South Africa}
\author{F.\ Montano}
\affiliation{Dipartimento di Fisica, Universit\`a degli Studi di Torino, Via P.\ Giuria 1, 10125 Torino, Italy}
\affiliation{INFN -- Istituto Nazionale di Fisica Nucleare, Sezione di Torino, Via P.\ Giuria 1, 10125 Torino, Italy}

\author{C.\ Clarkson}
\affiliation{Department of Physics \& Astronomy, Queen Mary University of London, London E1 4NS, United Kingdom}
\affiliation{Department of Physics \& Astronomy, University of the Western Cape, Cape Town 7535, South Africa}

\author{D.\ Karagiannis}
\affiliation{Dipartimento di Fisica e Scienze della Terra, Universit\`a degli Studi
di Ferrara, Via G.\ Saragat 1, 44122 Ferrara, Italy}
\affiliation{Department of Physics \& Astronomy, University of the Western Cape, Cape Town 7535, South Africa}

\author{R.\ Maartens}
\affiliation{Department of Physics \& Astronomy, University of the Western Cape, Cape Town 7535, South Africa}
\affiliation{National Institute for Theoretical and Computational Sciences, 
Cape Town 7535, South Africa}

\begin{abstract}
On the scale of the cosmic horizon, signatures that are unique to general relativity are concealed within the statistics of the large scale distribution of galaxies. These were thought to be beyond the reach of all but the most ambitious galaxy surveys, as they are substantially suppressed relative to standard redshift-space distortions. We show that the detectability of these higher-order relativistic effects can be dramatically enhanced by a sampling strategy that splits a galaxy catalogue into faint and bright subsamples and then combines their auto-bispectra. For current surveys such as DESI, this implies that this new signal will be detectable for the first time using our new strategy.
\end{abstract}

\maketitle

\paragraph*{\textbf{Introduction.}}
Spectroscopic galaxy surveys are beginning to access ultra-large scales, where relativistic light-cone projection effects imprint distinct signatures on the large-scale structure of the Universe. In particular, the galaxy three-point function in Fourier space, known as the bispectrum, has been shown to be more sensitive to such effects than the two-point function, or power spectrum \citep[e.g.][]{2017JCAP...03..034U,DiDio:2016gpd,2023PhRvD.107h3528N,Maartens:2019yhx,2019arXiv190605198J,2023JCAP...09..030P}. While subdominant on smaller scales, these relativistic corrections become increasingly important on super-equality scales and offer a potential test of general relativity on cosmological scales.

In \citet{Rossiter:2024tvi}, we forecast constraints on local primordial non-Gaussianity and relativistic effects using a galaxy bispectrum that includes corrections from both, modelled for the first time by \citet{Maartens:2020jzf}. We scrutinised the constraining power that upcoming galaxy surveys will have on detecting relativistic effects and the primordial non-Gaussianity amplitude parameter, \(\fnl\), with current surveys such as the Dark Energy Spectroscopic Instrument \citep[DESI,][]{2022AJ....164..207D,2025arXiv250314745D} and so-called Stage 5 spectroscopic galaxy surveys \citep[e.g. MegaMapper and the Wide-field Spectroscopic Telescope,][]{Schlegel:2022vrv,2024arXiv240305398M,2026arXiv260105320M}. We also showed that this formalism can lift degeneracies between non-Gaussianity and relativistic effects and we quantified that neglecting the latter in the galaxy bispectrum model can induce a bias of more than $2\,\sigma$ in $\fnl$ measurements.

Here, we focus on the capability of the bispectrum in detecting relativistic effects, looking at how to optimise sample selection in order to achieve the most significant detection. We follow \citet{2014PhRvD..89h3535B,2016JCAP...08..021B}, who proposed to split a parent galaxy sample into  faint and  bright subsamples and look at their two-point cross-correlation function, where a characteristic dipole appears uniquely due to relativistic effects \citep[see also][]{2024JCAP...12..029B}. This strategy has recently been further investigated in the galaxy power spectrum as well \citep{Montano:2023zhh,Montano:2024xrr,2025arXiv250908056N}, allowing the power spectrum to pick up imaginary signatures that would be otherwise cancelled out in the single-tracer case.

Furthermore, recent theoretical work by \citet{2025arXiv250418245R} strengthened the case for magnitude-based sample splitting: using an EFT-based Fisher forecast, they demonstrated that splitting by non-linear bias---naturally achieved via a magnitude cut---can significantly sharpen parameter constraints. In this communication we aim to scrutinise the sample-splitting strategy in the context of the relativistic galaxy bispectrum, focussing on an initial implementation that considers only the auto-bispectra of the resulting faint and bright galaxy subsamples.

Our case study is the DESI bright galaxy sample (BGS). Departing from the approach of \citet{Rossiter:2024tvi}, which obtains the survey parameters via a luminosity function, we now implement a halo occupation distribution framework for BGS galaxies \citep{2023AJ....165..253H}, as in \citet{Montano:2023zhh,Montano:2024xrr}. Thanks to this method, we can obtain a self-consistent treatment of the comoving number density, galaxy bias up to second order, and magnification and evolution biases as a function of both redshift and magnitude.
This also allows us to compute their derivatives with respect to redshift and magnitude, both of which enter the relativistic bispectrum \citep{Maartens:2020jzf}. We perform a tomographic analysis, assigning galaxies to redshift bins of width \(\Delta z=0.1\) in the range \(z\in[0.05,0.65]\).

\paragraph*{\textbf{Formalism.}} 
At tree level, the bispectrum of fluctuations in number counts of a tracer \(t\) is defined by (omitting explicit redshift dependence for clarity)
\begin{equation}
\Big\langle\Delta_t^{(1)}(\ka)\,\Delta_t^{(1)}(\kb)\,\Delta_t^{(2)}(\kc) \Big\rangle + 2\!\circlearrowleft = 2 \, (2\,\pi)^3\,B_t(\ka,\kb) \,\delta_{\rm D}(\kabc) \;,
\end{equation}
where $\circlearrowleft$ denotes cyclic permutations over $\ka$, $\kb$, and $\kc=-\kab$, $\bm k_{12\ldots N}$ is a shorthand notation for $\ka+\kb+\ldots+\bm k_N$, and we adopt the convention that the observed galaxy number density contrast is $\smash{\Delta_t = \Delta_t^{(1)} +\Delta_t^{(2)}/2}$, with superscript ${(m)}$ indicating the $m$th-order in the perturbed expansion. Hence, the galaxy bispectrum can be expressed in terms of Fourier kernels as
\begin{equation}
B_t(\ka, \kb)= \ko_t^{(1)}(\ka)\,\ko_t^{(1)}(\kb)\,\ko_t^{(2)}(\ka,\kb)\,P(k_1
)\,P(k_2)+2\!\circlearrowleft\;,\label{eq:bisp}
\end{equation}
where $P$ is the linear matter power spectrum and the kernels $\smash{\ko_t^{(m)}}=\ko^{(m)}_{t,{\rm N}} + \ko^{(m)}_{t,{\rm GR}}$ contain Newtonian and local (i.e.\ non-integrated) relativistic contributions. The full expressions for the kernels can be found in \citet{Maartens:2020jzf}. 

Relativistic contributions are naturally ordered not just in perturbative order but also in powers of Fourier wavelength compared to the conformal Hubble scale, $(\cH/k)^n$. Newtonian terms are all defined as $n=0$, and increasing $n$ implies contributions are in principle sub-dominant. At 1st-order in perturbation theory, contributions contain $n=1$ (Doppler and gravitational redshift) and $n=2$ (metric potentials, time delay, and Sachs-Wolfe-type terms). At 2nd-order, a much more complicated hierarchy emerges, with mixing between these effects and others, such as a transverse Doppler term \citep[see][]{2017JCAP...09..040J,Jolicoeur:2017eyi}. The bispectrum mixes these different contributions between perturbative orders, with $O(\cH/k)$ contributing to odd multipoles and $O(\cH^2/k^2)$ contributions (from both first and second order) being responsible for biasing $\fnl$ measurements.

Here, we focus on the optimisation of a flux cut to subdivide a galaxy catalogue (T, for `total') into a bright subsample (B) and a faint subsample (F), with the aim of maximising the detectability of relativistic effects. To assess the significance of the detectability, we construct a \(\Delta\chi^2\) variable as the difference between the \(\chi^2\) summary statistics of a model that contains all relativistic effects (as our Universe ought to be), and one---the null hypothesis---where the fit accounts for Newtonian terms only. Since we deal with noiseless synthetic data, in each redshift bin this simply corresponds to
\begin{equation}
\Delta\chi^2(z_i)\equiv\bm\beta_i^\dag\,{\sf C}_i^{-1}\,\bm\beta_i\;,
\end{equation}
in which \(\bm\beta_i\) and \({\sf C}_i\) are, respectively, the theory vector of pure relativistic contributions to the bispectrum and the covariance matrix, both evaluated in the \(i\)th redshift bin, and with \(\dagger\) denoting conjugate transposition.

The theory vector \(\bm\beta_i\) is the collection of the values of the relativistic bispectrum, \(B_{t,{\rm GR}}\coloneqq B_t-B_{t,{\rm N}}\), in all discretised triangle bins, where  \(B_{t,{\rm N}}\) is the outcome of \cref{eq:bisp} with only Newtonian kernels.

Regarding the covariance matrix, we work under the Gaussian assumption, in which it is diagonal in triangle shapes, with the per-triangle variance 
\begin{equation}
   \var\big[B_t(\ka, \kb)\big]=\frac{s_\triangle}{N_\triangle}\,\tilde P_t(\ka)\,\tilde P_t(\kb)\,\tilde P_t(-\kab)\;.\label{eq:var}
\end{equation}
Here \(s_\triangle\) is the multiplicity of a given triangle, being 6, 2, or 1 for equilateral, isosceles, or scalene bin configurations, respectively. Then \(N_\triangle\) is the number of triangles in a given triangle bin and \(\tilde P_t=P_t+1/n_t\), with \(\smash{P_t(\bm k) =  \ko_t^{(1)}(\bm k)\,\ko_t^{(1)}(-\bm k)\,P(k)}\) being the tree-level galaxy clustering power spectrum and \(n_t\) the volumetric number density of galaxies of tracer \(t\) (at the reference redshift).

In each redshift bin, we then define the signal-to-noise ratio (SNR) as \(\smash{\snr(z_i)=\sqrt{\Delta\chi^2(z_i)}}\), whereas the cumulative one is given by
\begin{equation}
\snr(\le z_i)^2={\sum_{j\le i}\snr^2(z_j)}\;,\label{eq:snr_cum}
\end{equation}
and the total SNR is obtained from \(\snr^2=\sum_i\snr^2(z_i)\).

\paragraph*{\textbf{Subsample optimisation.}}
\citet[][see also Appendix B in \citealp{Montano:2024xrr}]{Montano:2023zhh}  showed that the power spectrum of a subsample of a galaxy catalogue can yield better constraints on relativistic effects than when using the whole catalogue. This rather counter-intuitive fact stems from the peculiar sample-dependence of relativistic effects, due to the presence of the so-called magnification and evolution biases \citep[e.g.][]{Maartens:2021dqy}. The former is the slope of the logarithmic number counts at magnitude cut, whereas the latter accounts for the redshift evolution of number counts. More generally, the gain from sample splitting is controlled by specific combinations of tracers and can vary strongly with selecion criteria \citep{Barreira:2023rxn}.

Specifically, by cutting at a brighter magnitude, we select a bright subsample with a different magnification and evolution bias (as well as a larger clustering bias); and, possibly, a different number density too. This in turn reflects onto the complement set, i.e.\ the faint subsample, since all the biases can be written as sums weighted by the number densities. That is, \(X_{\rm T}\,n_{\rm T}=X_{\rm F}\,n_{\rm F}+X_{\rm B}\,n_{\rm B}\), where \(X\) can be any among the linear bias, magnification bias, and evolution bias. Hence, we can optimise the splitting magnitude \(\ms\) between the faint and the bright subsample, looking for the combination of biases that effectively boosts the relativistic contributions, maximising \(\smash{\ko^{(i)}_{t{\rm GR}}}\).

\begin{figure}
    \centering
    \includegraphics[width=\columnwidth]{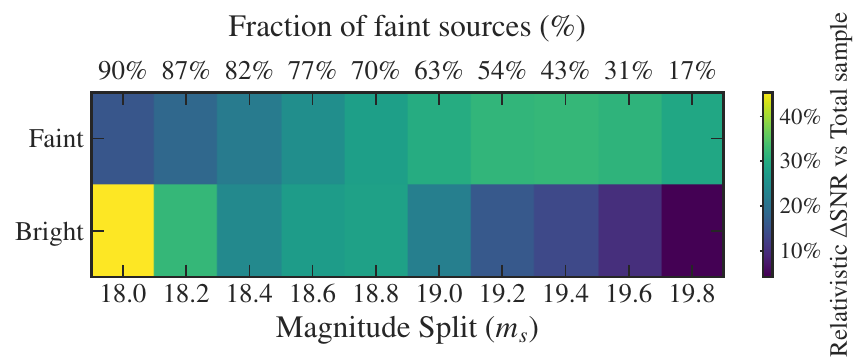}
    \caption{Percentage difference of SNRs for the non-Newtonian faint- and bright-subsample bispectra compared to the bispectrum of the total sample, across different magnitude splits, $m_{\rm s}\in[18,19.8]$.   }
    \label{fig:FBheatmap}
\end{figure}

Here, we attempt for the first time such an optimisation for the bispectrum. \Cref{fig:FBheatmap} shows the results of the optimisation process. The heatmap refers to percentage change in the SNR w.r.t.\ using the bispectrum of the total catalogue, as a function of \(\ms\). As a guidance to the eye, note that the closer \(\ms\) is to the total sample's magnitude cut of \(\mc=20\), the more the faint subsample is depopulated, and the bright subsample effectively becomes the total sample. (Indeed, the relative change in the SNR in the second row tends to \(0\%\) as we move to the right.) In general, we see that the faint bispectrum yields larger SNR improvements---a trend also observed in the power spectrum \citep{Montano:2023zhh,2023MNRAS.525.4611B}. Most interestingly, there is a sweet spot at which the increment reaches a maximum, around \(\ms\simeq19.3\). In this case, the SNR for the faint bispectrum is more than \(30\%\) larger than for the total sample, which in the present case of DESI BGS changes a non-detection ($\rm {SNR}\approx 2.5$) to a detection ($\rm {SNR}\approx 4$). 

\paragraph*{\textbf{Joint analysis of faint- and bright-subsample bispectra.}}
Given that both subsamples provide a larger SNR than the total sample, it is worth asking what results can be attained by using them jointly. Thus, for a combination of the faint-only and bright-only bispectra, in each redshift bin we construct a joint data vector
\begin{equation}
   \bm\beta=
    \begin{pmatrix}
    B_{\rm F,GR} &
    B_{\rm B,GR}
    \end{pmatrix}^\dagger
\end{equation}
with a covariance matrix
\begin{equation}
{\sf C}=
    \begin{pmatrix}
        \var\left[B_{\rm F}\right]& \cov\left[B_{\rm F},B_{\rm B}\right]\\
        \cov\left[B_{\rm B},B_{\rm F}\right] & \var\left[B_{\rm B}\right]
    \end{pmatrix}\;,\label{eq:cov_mat}
\end{equation}
where
\begin{equation}
    \cov\left[B_t(\ka, \kb),B_{t'}(\ka, \kb)\right]=\frac{s_\triangle}{N_\triangle}\,\tilde P_{tt'}(\ka)\,\tilde P_{tt'}(\kb)\,\tilde P_{tt'}(-\kab)\label{eq:cov}
\end{equation}
is the covariance between the bispectrum of tracer \(t\) and that of tracer \(t'\) \citep{Karagiannis:2023lsj}. Here, \(\smash{P_{tt'}(\bm k) =  \ko_t^{(1)}(\bm k)\,\ko_{t'}^{(1)}(-\bm k)\,P(k)}\) is the tree-level cross-spectrum between the two tracers, and as long as the two tracers are disjoint galaxy populations, \(\tilde P_{tt'}\equiv P_{tt'}\). Note that, since the relativistic bispectrum is a complex number, the covariance matrix \({\sf C}\) has to be Hermitian, implying \(\cov[B_{\rm B},B_{\rm F}]=\cov[B_{\rm F},B_{\rm B}]^\ast\).

\begin{figure}
    \centering
    \includegraphics[width=\columnwidth]{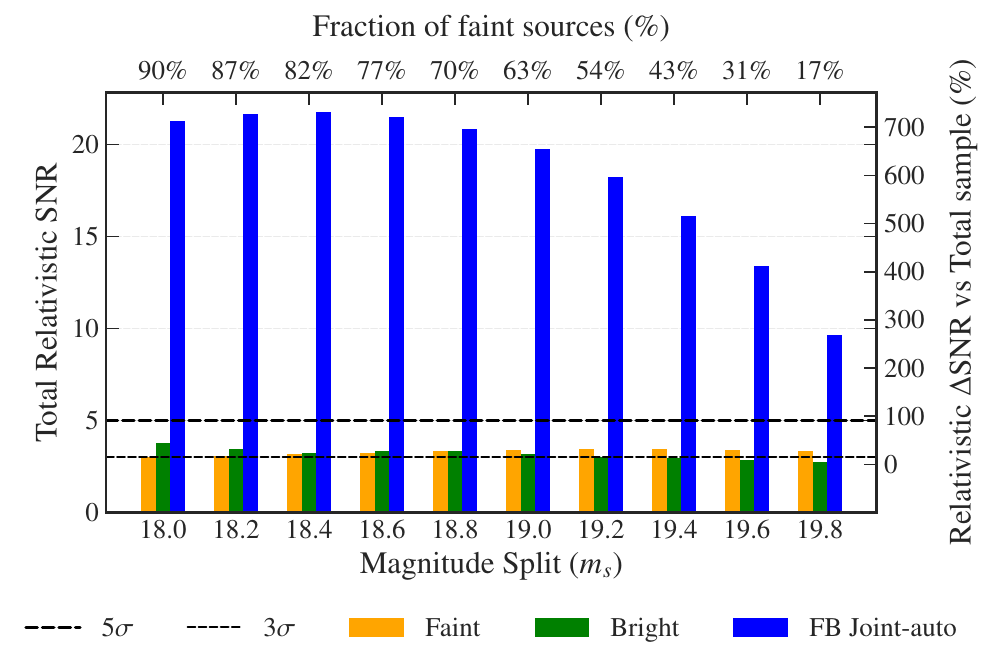}
    \caption{Histogram demonstrating the percentage increase of the total relativistic SNR for faint- and bright-subsamples and the F+B  bispectra compared to the bispectrum of the total sample, across different magnitude splits, $m_{\rm s}\in[18,19.8]$. Dashed horizontal lines show the $5$ and $3\,\sigma$ detection significance thresholds.}
    \label{fig:F+Bhistogram}
\end{figure}

We compute the SNR for the relativistic contribution to the bispectrum in the joint faint and bright subsample analysis across a range of splitting magnitudes $\ms$ in order to find an optimum. The result is shown in \cref{fig:F+Bhistogram}, where ordinates on the left axis show the value of the total SNR for the joint F+B bispectrum (blue bars) compared to the faint and bright subsamples separately (yellow and green bars, respectively). The corresponding increase with respect to the total sample is shown by ordinates on the right axis. We can appreciate a clear trend, and a peak resulting in a \(\snr=22.5\) is found at \(\ms=18.4\), corresponding roughly to a \(80\)--\(20\) split in number between faint and bright galaxies.

The total signal is made up of the sum of triplets of first- and second- order kernels as per \cref{eq:bisp}. The relativistic signal can be thought of as the purely Newtonian bispectrum subtracted from the total bispectrum. This results in triplets that contain purely relativistic combinations of kernels, \(\ko_{t,{\rm GR}}^{(1)}\,\ko_{t,{\rm GR}}^{(1)}\,\ko_{t,{\rm GR}}^{(2)}\), but also triplets factoring relativistic and Newtonian kernels, e.g.\ \(\ko_{t,{\rm N}}^{(1)}\,\ko_{t,{\rm GR}}^{(1)}\,\ko_{t,{\rm GR}}^{(2)}\). The leading contribution to the relativistic kernel is the Doppler term, which enters as $\mathcal{H}/k$ and contributes an imaginary signal to the bispectrum. These couplings between first order Doppler terms and first and second order Newtonian kernels are where the signal is boosted most significantly.

\begin{figure}
    \centering
    \includegraphics[width=\columnwidth]{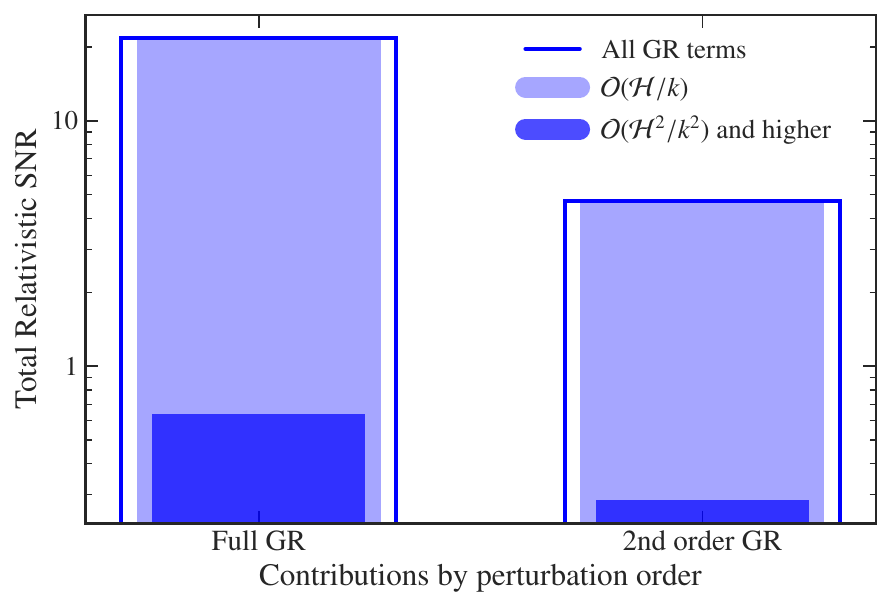}
    \caption{Comparison of relativistic contributions to the relativistic bispectrum SNR, full GR and 2nd-order only. Shading corresponds to contributions grouped by scaling with powers of $\cH/k$.}
    \label{fig:SNR_contributions}
\end{figure}

We look to \cref{fig:SNR_contributions} to verify this, where the relativistic SNR is computed at the optimal magnitude split discussed above. Here we can compare contributions from different \((\cH/k)^n\) powers and perturbative orders to the relativistic bispectrum SNR. We can easily see that the majority of the relativistic signal comes from \(n=1\) terms. As a comparison, this also applies in the power spectrum, although none of the contributions reach an $\snr=1$, but with the crucial difference that $n=2$ corrections counteract---rather than amplify---the dominant ones. That being said, the use of sample splitting techniques makes even \(n\ge2\) terms in the bispectrum more significant than in a standard analysis.

An important aspect of \cref{fig:SNR_contributions}, appreciable by comparing the right bar to the left one, is that 2nd-order relativistic terms are measurable and contribute a substantial fraction of the full signal. This means that they need to be modelled properly and included in a full analysis. This also presents the new possibility of using these extra subtle relativistic effects as tests of general relativity in future.  However we can also see that we do not need to go far in the $(\mathcal{H}/k)$ hierarchy to saturate the modelling accuracy for this sample at least.

\begin{figure}
    \centering
    \includegraphics[width=\columnwidth]{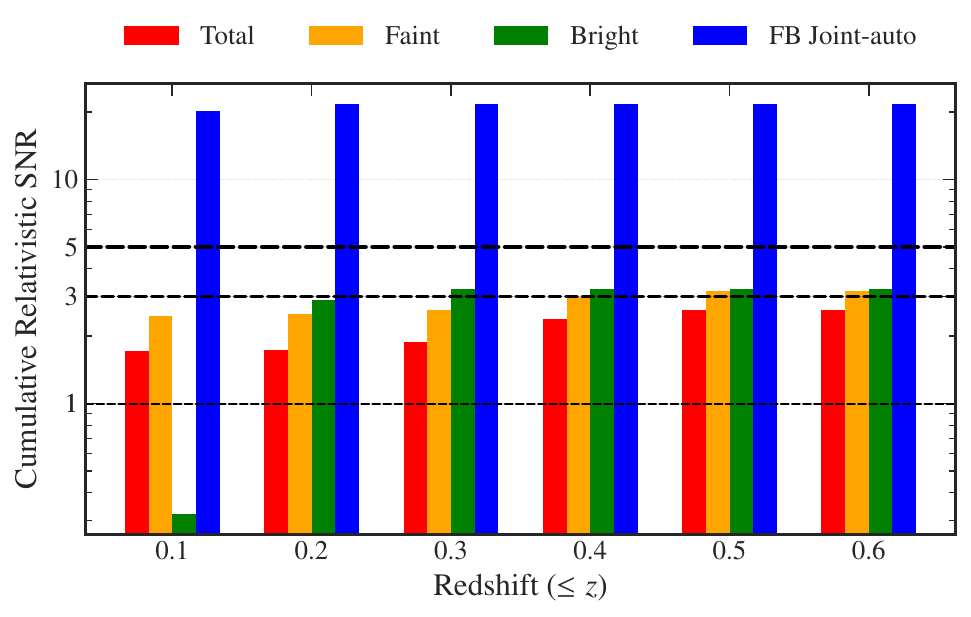}
    \caption{Comparison of the cumulative relativistic SNR for each sample analysis at the optimum magnitude split of $m_{\rm s}=18.4$ ($82 \%$ faint sources).}
    \label{fig:snr_samples_cum}
\end{figure}

This can be seen also in the cumulative SNR, shown in \cref{fig:snr_samples_cum}, for the total bispectrum (red), the faint bispectrum (yellow), the bright bispectrum (green) and the faint-bright joint analysis (blue) at the optimum splitting magnitude, \(\ms=18.4\). Dashed lines show the notable detection significance thresholds.
The improvement over the standard case is staggering, amounting to an SNR as much as \({\sim}\,9\) times larger. This makes a signal that appears to be barely measurable in the DESI BGS ripe for a potentially high-significance detection, even when real-world effects are taken properly into account.

It is also instructive to note how the SNR values basically saturate in the first redshift bin for all but the bright subsample. This confirms our expectations that the Doppler contribution is more dominant at low redshifts where peculiar velocities are substantial and the conformal Hubble parameter $\mathcal{H}$ is smaller.

It is both the gain in constraining power, which comes from the joint faint-bright analysis, and the high sensitivity of the bispectrum to GR corrections that allows us to reach such a high SNR in \cref{fig:snr_samples_cum}. We cannot attain the same without one of the two. Indeed, an analogous study conducted on auto-power spectra---simpler to be dealt with, yet not as suited for seeking for relativistic effects as bispectra are---yields a factor of ${\sim}\,5(4)$ improvement in the SNR of the joint measurement, w.r.t.\ the bright(faint) subsample alone, though with an overall $\rm{SNR}<1$.

\paragraph*{\textbf{Conclusions.}}
We have demonstrated that the relativistic contributions to the galaxy bispectrum, dominated by the Doppler term, can be detected in current surveys once an optimal faint/bright split is applied. We present DESI BGS as a prime candidate for this application, where the total sample alone would yield a marginal \(\snr \approx 2.5\), but subdividing the sample boosts this to \(\snr \approx 22.5\), an improvement of an order of magnitude. This enhancement arises primarily from 1st-order GR Doppler terms coupling with standard Newtonian terms, while higher-order Doppler couplings play a subdominant yet non-negligible role in shaping the signal. Importantly, we find that even the faint-only or bright-only analyses outperform the total sample, offering practical advantages for smaller datasets with reduced covariance size and simulation demands.

\begin{acknowledgments}
SJR, FM, and SC acknowledge support from the Italian Ministry of University and Research (\textsc{mur}), PRIN 2022 `EXSKALIBUR – Euclid-Cross-SKA: Likelihood Inference Building for Universe's Research', Grant No.\ 20222BBYB9, CUP D53D2300252 0006, from the Italian Ministry of Foreign Affairs and International Cooperation (\textsc{maeci}), Grant No.\ ZA23GR03. DK acknowledges support by \textsc{mur}, PRIN 2022 `BROWSEPOL: Beyond standaRd mOdel With coSmic microwavE background POLarization', Grant No.\ 2022EJNZ53. SJR, FM, SC, and DK acknowledge support from the European Union -- Next Generation EU. CC is supported by STFC grant ST/X000931/1. RM was supported by the South African Radio Astronomy Observatory and National Research Foundation, Grant No. 75415.
\end{acknowledgments}

\bibliography{biblio.bib}

\end{document}